\documentclass[10pt,aps.prl,twocolumn,superscriptaddress,floatfix]{revtex4-1} 

\usepackage{amsmath}
\usepackage{graphics}
\usepackage{color}
\usepackage{caption}
\usepackage{subcaption}
\usepackage{float}
\usepackage{array}
\usepackage[margin=1.0in]{geometry}
\usepackage{ulem}
\usepackage{amssymb}

\begin{document}
 %revtex intro:
 \title{Photodegradation and Self-Healing in a Rhodamine 6G Dye and Y$_2$O$_3$ Nanoparticle-Doped Polyurethane Random Laser}
\author{Benjamin R. Anderson$^1$, Ray Gunawidjaja$^1$, and Hergen Eilers$^{*}$}
\affiliation{Applied Sciences Laboratory, Institute for Shock Physics, Washington State University,
Spokane, WA 99210-1495}
\date{\today}

\email{eilers@wsu.edu}
 
%APB intro:
% \title{Photodegradation and Self-Healing in a Rhodamine 6G Dye and Y$_2$O$_3$ Nanoparticle-Doped Polyurethane Random Laser}
% \author{Benjamin Anderson \and Ray Gunawidjaja \and Hergen Eilers}
% \institute{Applied Sciences Laboratory, Institute for Shock Physics, Washington State University,
% Spokane, WA 99210-1495}
% \mail{eilers@wsu.edu}
% \date{\today}

\begin{abstract}
One of the fundamental difficulties in implementing organic dyes in random lasers is irreversible photodegradation of the dye molecules, leading to loss of performance and the need to replace the dye. We report the observation of self-healing after photodegradation in a Rhodamine 6G dye and nanoparticle doped polyurethane random laser.  During irradiation we observe two distinct temporal regions in which the random lasing (RL) emission first increases in intensity and redshifts, followed by further redshifting, spectral broadening, and decay in the emission intensity.  After irradiation the emission intensity is found to recover back to its peak value, while still being broadened and redshifted, which leads to the result of an enhancement of the spectrally integrated intensity.  We also perform IR-VIS absorbance measurements and find that the results suggest that during irradiation some of the dye molecules form dimers and trimers and that the polymer host is irreversibly damaged by photooxidation and 
Norrish type I photocleavage.
\vspace{1em}

\end{abstract}

\maketitle

\vspace{1em}

\section{Introduction}

% \footnotetext{\textit{$^{a}$~Address: Applied Sciences Laboratory, Institute for Shock Physics, Washington State University,
% Spokane, WA 99210-1495. Tel: (509)358-7681; E-mail: eilers@wsu.edu}}

Lasing in disordered media was first predicted in the late 1960's by Letokhov \cite{Letokhov66.01,Letokhov67.01,Letokhov67.02} and first demonstrated in the mid 1990's \cite{Lawandy94.01}.  This form of lasing -- in which scattering acts as the feedback mechanism -- is known as random lasing.  Random lasing (RL) is found to have two regimes of behavior: intensity feedback random lasing (IFRL) and resonant feedback random lasing (RFRL) \cite{Cao03.01,Cao03.02,Cao05.01}.  IFRL is characterized by a single emission peak with a linewidth of approximately 5-10 nm \cite{Cao03.02,Lawandy94.01,Cao05.01} while RFRL is characterized by the appearance of multiple sub-nm peaks \cite{Cao03.02,Meng09.01,Cao99.01,Ling01.01,Cao05.02}.  The sharp peaks of RFRL arise due to a small number of lasing modes in the scattering material being active. As the number of active modes increases they merge into a broad peak corresponding to IFRL \cite{Cao03.01,Cao03.02,Ling01.01,Tureci08.01,Ignesti13.01}.

The multimode nature of IFRL results in the emission having a low degree of spatial coherence \cite{Redding11.01}, which makes it attractive as a high-intensity low-coherence light source \cite{Redding12.01}, which has applications in biological imaging \cite{Redding12.01,Hecht12.01}, picoprojectors, cinema projectors \cite{Hecht12.01}, photodynamic therapy, tumor detection \cite{Hoang10.01,Cao05.01}, flexible displays, and active elements in photonics devices \cite{Cao05.01}. On the other hand, lasing in few modes creates unique spectral signatures that can be used in encoding for authentication, labeling of cells for biological imaging, and emergency beacons \cite{Hoang10.01,Cao05.01}.  Additionally, the spectrum of a RFRL can be controlled by spatial modulation of the pump beam \cite{Leonetti12.01,Cao05.01,Leonetti13.02,Leonetti12.02,Leonetti12.03,Andreasen14.01,Bachelard12.01,Bachelard14.01}. Controlled RFRL can be used to create tunable light sources \cite{Cao05.01} and implement optically based 
physically unclonable functions \cite{Anderson14.04,Anderson14.05}.

While RL has been observed in a wide variety of materials, one of the primary material classes used for studying RL is organic dyes and nanoparticles doped into solution or polymer.  These materials are attractive as they are easy and cheap to produce, and allow for the formation of arbitrary shapes \cite{Cao05.01}.  While organic dyes have these attractive features, they also have a major flaw in that they tend to irreversibly photodegrade under extended exposure to high intensity light \cite{wood03.01,taylo05.01,Avnir84.01,Knobbe90.01,Kaminow72.01,Rabek89.01}.  Irreversible photodegradation leads to a loss in performance and efficiency, which eventually requires the dye to be replaced if the random laser is to continue to work.

A possible solution to irreversible photodegradation of organic-dye-based random lasers is to make random lasers using self-healing dye-doped polymers.  Self-healing dye-doped polymers are materials which display the characteristics of irreversible photodegradation when exposed to intense light, but once the light is removed the material ``heals'' itself returning to its original efficiency.  The phenomenon of self-healing in dye-doped polymers was first observed by Peng \textit{et al.} in Rhodamine B and Pyrromethene dye-doped poly(methyl-methacrylate) (PMMA) optical fibers \cite{Peng98.01}. Since Peng \textit{et al.'s} observation, self-healing has been observed in disperse orange 11 (DO11) dye-doped PMMA \cite{howel04.01,howel02.01}, DO11 dye-doped copolymer of styrene and MMA \cite{Hung12.01}, air force 455 (AF455) dye-doped PMMA \cite{Zhu07.01}, 8-hydroxyquinoline (Alq) dye-doped PMMA \cite{Kobrin04.01}, anthraquinone-derivative-doped PMMA \cite{Anderson11.02}, and Rhodamine 6G dye-doped polyurethane 
with dispersed ZrO$_2$ nanoparticles \cite{Anderson15.01}.

In this study we present the observation of self-healing in random lasers composed of Rhodamine 6G dye-doped polyurethane with dispersed Y$_2$O$_3$ nanoparticles.  We find that photodegradation results in a reversible decrease in the peak lasing intensity, while also resulting in irreversible changes to the lasing wavelength and linewidth.  We also perform IR-VIS absorbance measurements to determine probable mechanisms of the irreversible changes in the random laser due to photodegradation.

\section{Method}
\begin{figure}
 \centering
 \includegraphics{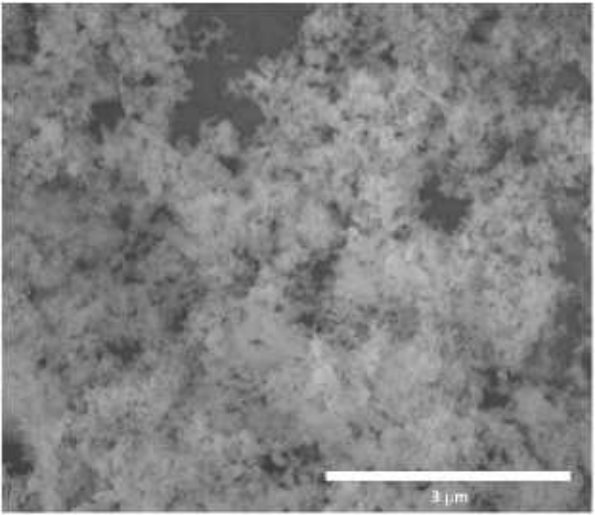}
 \caption{SEM image of Y$_2$O$_3$ NPs. The NPs are found to form agglomerates with a size on the order of microns.}
 \label{fig:SEM}
\end{figure}

The random laser samples used for this study are prepared as follows.  We first dissolve Rhodamine 6G (R6G) in tetraethylene glycol (TEG) such that the resulting dye concentration is 1 wt\% of the polymer.  To prepare the polymer we mix equimolar amounts of TEG and poly(hexamethylene diisocyanate) (pHDMI) together along with 0.1 wt\% di-n-butyltin dilaurate (DBTDL) catalyst, where the TEG, pHMDI, and DBTDL are purchased from Sigma-Aldrich.  The resulting dye-doped mixture is then centrifuged for 3 min at 3000 rpm to remove air bubbles, at which point it is poured into a 1" circular die.  Next we add a dispersion of Y$_2$O$_3$ NPs (Inframat, 30-50 nm diameter, Figure \ref{fig:SEM} shows an SEM image) in 1,4-dioxane and DBTDL to the mixture and stir until it is homogeneous.  The mixture is then left at room temperature to cure over night.

\begin{figure}
\centering
\includegraphics{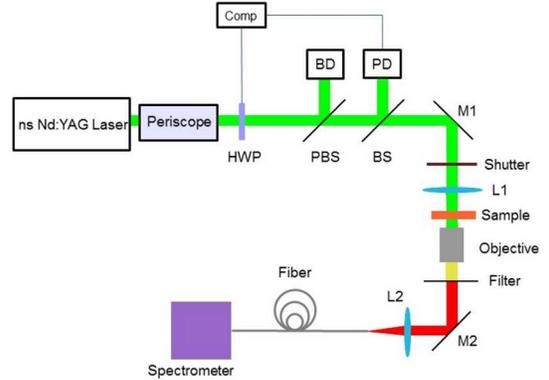}
\caption{Schematic of intensity stabilized random lasing experiment.  A beam sampler (BS) picks off a portion of the pump beam for a photodiode (PD) to provide a feedback signal to a half-waveplate (HWP)/polarizing beamsplitter (PBS) combination.}
\label{fig:schematic}
\end{figure}

Photodegradation and recovery of the R6G+ Y$_2$O$_3$/PU samples is measured using an intensity stabilized random lasing system consisting of a Spectra-Physics Quanta-Ray Pro frequency doubled Nd:YAG laser (532 nm, 10 ns, 10 Hz), a motorized half-waveplate (HWP), polarizing beam splitter (PBS), a Thorlabs Si photodiode, a motorized shutter, a plano-convex lens ($f = 50$ mm), and a Princeton Instruments spectrometer.  See Figure \ref{fig:schematic} for a schematic of the experimental setup. 

The laser's power is monitored and controlled using a feedback loop between the photodiode and motorized half-waveplate, with five pulse averaging used to determine the feedback signal.  Using this technique we maintain the on sample intensity within 1\%  over the entire experimental time period. The main limitation to reducing the intensity variance further is the laser's shot-to-shot noise value of approximately 1\%. For our various trials we use pump energies between 1.5 mJ and 33 mJ with an on sample spot size of 2.8 mm, giving an intensity range of 2.44 MW/cm$^2$ to 53.6 MW/cm$^2$. 

Additionally, we perform imaging and IR-VIS absorbance measurements to characterize the underlying changes in the material due to irradiation.  For VIS absorbance measurements we focus white light to a 0.25 mm spot on the sample and collect the reflected light into a spectrometer.  The reason for using reflected light is that the majority of the photo-induced changes occur near the surface of the sample and, due to scattering, reflected light is a more accurate probe of surface changes than transmitted light, which undergoes more scattering and a longer optical pathlength in the sample.  

For these reasons our FTIR measurements are also performed in reflection using an FTIR-ATR (Varian 680) using a frequency range of 4000 cm$^{-1}$ to 400 cm$^{-1}$, a step size of 1 cm$^{-1}$, and 32 spectrum averaging. To better differentiate between photo-induced changes in the IR spectrum and those due to variations within the sample, we measure the pristine IR absorbance at five different points on the sample and compute the average absorbance and standard deviation.  We find that the pristine sample has a consistent IR absorbance at all points tested and that changes in the IR absorbance spectrum due to irradiation are larger than the point-to-point variations.

\section{Results and Discussion}
\begin{figure}
\centering
\includegraphics{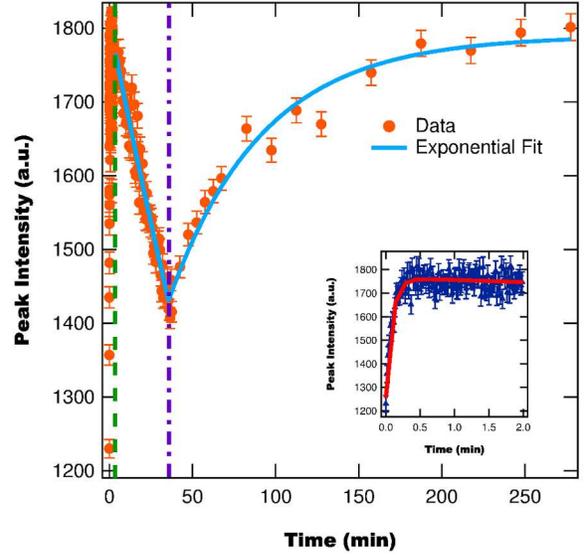}
\caption{Peak intensity as a function of time during conditioning (0 min $\leq t \leq$ 0.2 min), degradation (0.2 min $< t \leq$ 37 min), and recovery ($t > 37 $min).  The ``end'' of conditioning is marked by the vertical dashed line and the end of degradation is marked by the vertical dash-dot line.  The decay is fit to a single exponential function with a rate of  $8.15(\pm 0.27)\times10^{-3}$ min$^{-1}$ and the recovery is fit to a single exponential function with a recovery rate of $\beta =1.61(\pm 0.26)\times 10^{-2}$ min $^{-1}$.  The inset shows the intensity during and immediately after conditioning.}
\label{fig:amp}
\end{figure}

\begin{figure}
\centering
\includegraphics{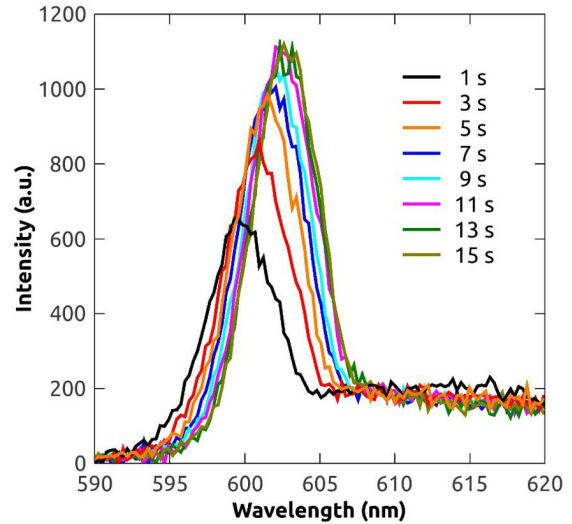}
\caption{Random lasing spectra at several times during conditioning.  The peak is found to red shift and increase in amplitude.}
\label{fig:condspec}
\end{figure}

We begin our study of photodegradation and self-healing in R6G+Y$_2$O$_3$/PU random lasers by considering the peak lasing intensity as a function of time during pumping and recovery, shown in Figure \ref{fig:amp}.  From the figure we observe three distinct temporal regions. The first region, which we call conditioning,  begins the moment the pump laser is incident on the material and lasts for $\approx$ 0.2 min (120 pulses).  This region is characterized by a rapid increase in peak intensity and a large red-shift in the emission spectrum. Figure \ref{fig:condspec} shows spectra at several times during conditioning. While we observe conditioning of the RL spectra in R6G+ Y$_2$O$_3$/PU, this effect is not observed in R6G+ZrO$_2$ /PU \cite{Anderson15.01}, suggesting that the phenomenon is related to the properties of the NPs dispersed into the polymer.  Most likely the different NPs affect the polymerization of the sample as FTIR results, discussed below, show that conditioning corresponds to photothermal 
induced polymerization of leftover monomer in the system. 

After conditioning, the random laser emission decreases in intensity, red-shifts, and is broadened.  After degradation the pump beam is blocked by a shutter and the sample begins to self-heal. During recovery, at semi-log time intervals, the pump shutter is opened for three pulses and the RL emission is measured.  As the sample self-heals the RL emission intensity is found to increase back to its conditioned value (see Figure \ref{fig:amp}).

\begin{figure}
\centering
\includegraphics{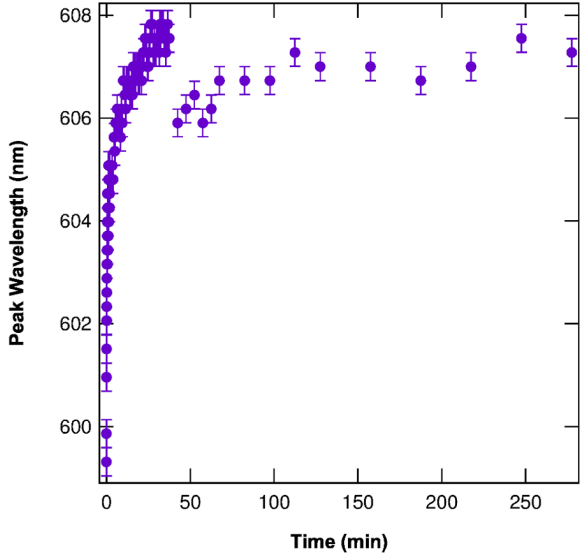}
\caption{RL peak wavelength as a function of time during conditioning, photodegradation, and recovery.  The peak wavelength is red-shifted during conditioning and photodegradation followed by a sharp blue-shift once the laser is blocked.}
\label{fig:center}
\end{figure}

While the peak intensity is found to return to it's conditioned value, the emission wavelength and linewidth are found to irreversibly change due to photodegradation.  Figure \ref{fig:center} shows the peak wavelength as a function of time, where it rapidly redshifts during conditioning and degradation, but then blueshifts when the pump beam is blocked and the sample is left in the dark for 15 min.  From the observation of a blue shift when the laser is turned off we hypothesize that part of the red shift is due to photothermal heating \cite{Anderson15.01}, as an increase in temperature is known to result in a red shift of RL emission \cite{Vutha06.01}.  To address the remaining redshift we propose that the polymer undergoes irreversible photo-induced changes, which modify the index mismatch of the system \cite{Anderson14.02} resulting in an increase in scattering. This hypothesis  is supported by the observation that increases in the amount of scattering in a material results in the lasing wavelength 
becoming redshifted \cite{Yi12.01,Shuzhen09.01,Anderson14.04,Dong11.01}.  Later in the paper we will identify several possible candidates for the photo-induced modifications of the polymer host.

\begin{figure}
\centering
\includegraphics{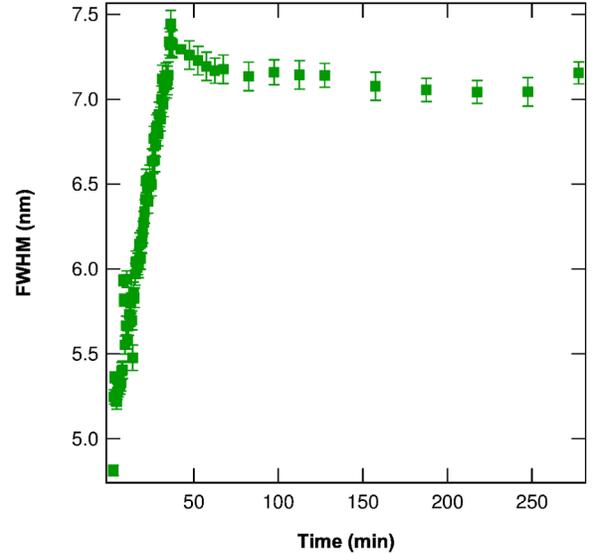}
\caption{Linewidth of RL as a function time during photodegradation and recovery.  During photodegradation the linewidth increases from 4.8 nm to 7.4 nm and then decreases to 7.1 nm during recovery.}
\label{fig:width}
\end{figure}

Along with the lasing wavelength redshifting during photodegradation, we also observe the lasing linewidth increasing, as shown in Figure \ref{fig:width}.  The FWHM is found to increase from 4.8 nm to 7.4 nm during degradation, and then to decrease slightly to 7.1 nm during recovery.  The increase in linewidth suggests the addition of new lasing modes to the lasing emission, as the linewidth of IFRL depends on the number of active modes \cite{Cao03.02,Cao03.01,Ling01.01,Tureci08.01,Ignesti13.01,Redding11.01}.  These new modes most likely arise due to new scattering resonances formed when the polymer is damaged and its refractive index is modified.  The slight decrease during recovery can be explained by slow chemical/mechanical changes in the polymer after illumination is removed.

While the general trends of the linewidth and peak wavelength during photodegradation and recovery are consistent for both R6G+Y$_2$O$_3$/PU and a previous study on R6G+ZrO$_2$/PU \cite{Anderson15.01}, we note that the magnitudes of the effects are quite different. Namely, for the ZrO$_2$ NPs the emission experiences a redshift of approximately 3 nm and a linewidth increase of 0.8 nm, while for the Y$_2$O$_3$ NPs the emission experiences a redshift of 9 nm and a linewidth increase of 2.8 nm. In both cases the pump fluences are similar, suggesting that the mechanism behind the different magnitudes is related to the NPs dispersed in the two systems. One possible explanation is that the different sizes of the NPs (average diameter of ZrO$_2$ NPs is 250 nm, while the average diameter of the Y$_2$O$_3$ NPs is 40 nm) influences the irreversible photodegradation of the polymer, which is the mechanism behind the wavelength's redshift and linewidth's broadening.

\begin{figure}
 \centering
 \includegraphics{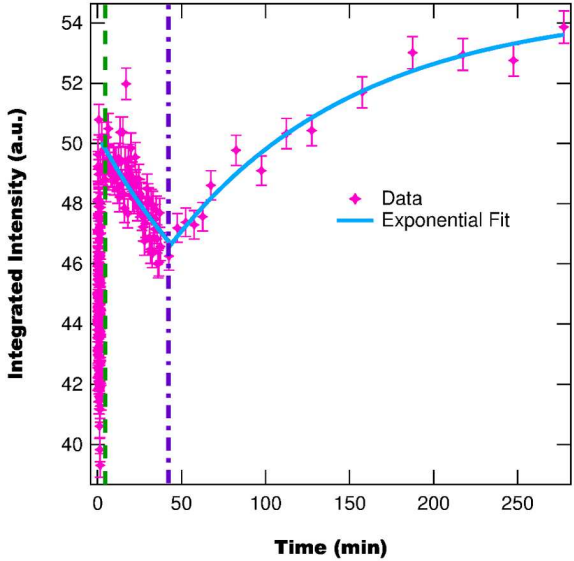}
 \caption{Integrated intensity as a function of time during decay and recovery. After recovery the integrated intensity is found to be greater than before decay by a factor of $1.085\pm 0.017$.}
 \label{fig:intI}
\end{figure}

Due to the peak intensity fully recovering and the linewidth increasing, we find that the spectrally integrated intensity increases after a decay and recovery cycle.  Figure \ref{fig:intI} shows the integrated intensity as a function of time, with it recovering to a higher value than before degradation. Fitting the recovery curve to an exponential curve gives an asymptotic intensity that is $1.085 \pm 0.017$ times greater than the conditioned sample's emission intensity. The observation of an enhanced integrated intensity due to photodegradation and recovery is consistent with previous measurements in R6G+ZrO$_2$/PU \cite{Anderson15.01}.

The enhancement of the broadband IFRL intensity of R6G+NP/PU is of special importance to applications requiring high-intensity low-coherence light, such as optical coherence tomography \cite{Redding11.01,Redding12.01} and laser based projectors \cite{Hecht12.01} where scattering results in unwanted interference effects. Previously, it was demonstrated that broadening the RL emission of a system results in the light becoming less coherent, thereby minimizing the effects of interference due to scattering \cite{Redding12.01}. This implies that by increasing the linewidth of R6G+NP/PU's RL emission -- while maintaining peak intensity -- photodegradation and recovery cycling results in the material having better broadband performance than the pristine material.

\subsection{Photodegradation and recovery cycling}
\begin{figure}
\centering
\includegraphics{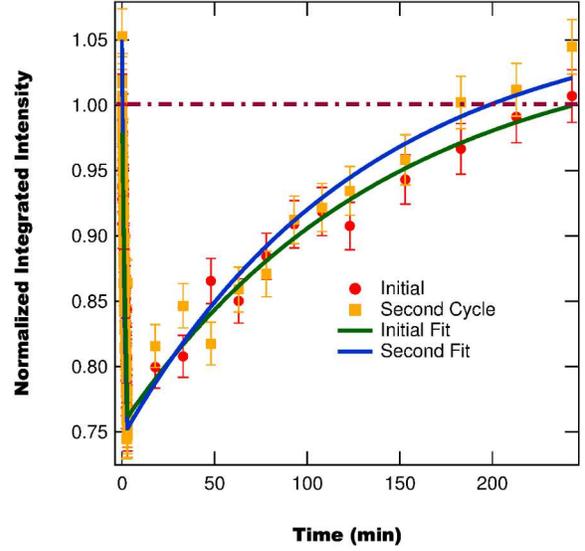}
\caption{Integrated intensity (normalized to the conditioned value) as a function of time during the initial and secondary decay and recovery cycle.}
\label{fig:2cint}
\end{figure}

Based on the observation of an enhancement in the integrated intensity as a result of photodegradation and recovery, we perform measurements in which we put the material through two cycles of degradation and recovery to see how much enhancement can be achieved. Both cycles are performed using the same intensity and timing to make direct comparisons. Figure \ref{fig:2cint} shows the integrated intensity during decay and recovery for both cycles. During the first cycle the integrated intensity decays 29\% at a rate of $0.125 \pm 0.025$ min$^{-1}$, after which it recovers to a value $1.064 \pm  0.015$ times larger than the conditioned integrated intensity at a rate of $\beta =7.86(\pm 0.79)\times 10^{-3}$ min $^{-1}$, which is within experimental uncertainty of the recovery rate determined for R6G+ZrO$_2$/PU \cite{Anderson15.01}.  After a period of twenty hours we cycle the sample once again and find that the integrated intensity now decays 37\% at a rate of $0.667 \pm 0.048$ min$^{-1}$, which is both more rapid and to a higher degree than the first cycle.  After the second degradation the integrated intensity recovers at 
the same rate as the first cycle, but now to a value $1.071 \pm 0.020$ times larger than the conditioned integrated intensity, which is within uncertainty of the first cycle's enhancement. 
%Peak
% DecA=29.32%  DecB=36.589%, decA rate = 0.11  decB rate= 0.702 \pm 0.078

%Integrated
%First cycle decayed 29% with a rate of  0.125 \pm 0.025 min^-1....second decayed 37% with a rate of 0.667 \pm 0.048  min^-1, data from 11/12/14
%First cycle gives integrated enhancement of 7%, second gives enhancement of 1%.......1.064 \pm  0.015  and 1.011  \pm 0.017

From these results we see that while the first decay and recovery cycle results in an enhancement of the integrated intensity, additional cycles contribute little, if any, additional enhancement.  Additional measurements using different amounts of decay in the first degradation cycle support this observation.  However, they also reveal that the amount of enhancement in the second cycle depends strongly on the amount of decay in the first cycle. Strong first decays result in little enhancement in the second decay, while light first decays result in larger enhancements in the second decay. These results suggest that continued cycling of the material results in the broadening linewidth reaching a steady state value, which stops further enhancement of the integrated intensity.

\subsection{Lasing Threshold of the Recovered System}
Thus far we have considered the emission properties of the system at a fixed pump intensity. Next we consider the peak intensity and linewidth as a function of pump energy in order to compare the lasing threshold of the conditioned and recovered sample.

\begin{figure}
\centering
\includegraphics{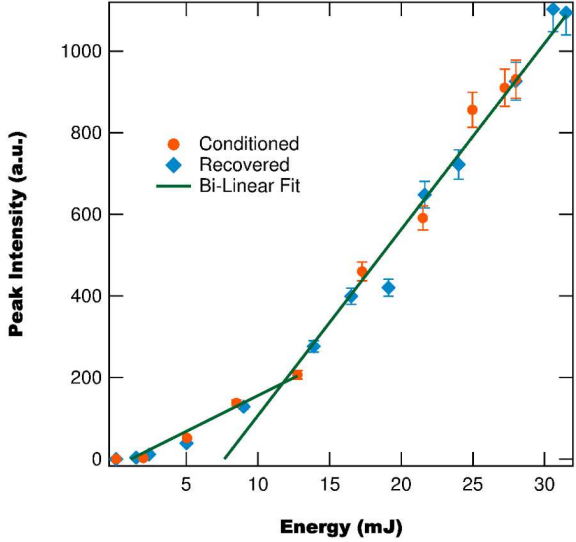}
\caption{Peak intensity as a function of pump energy for the conditioned sample (squares) and the sample after recovery (circles) and a bi-linear fit used to determine the lasing thresholds.  The intensities of the conditioned and recovered samples are found to be within uncertainty of each other, which is consistent with a recovery efficiency of 100\%.}
\label{fig:thresh}
\end{figure}

We first measure the peak intensity as a function of pump energy for the conditioned and recovered sample, shown in Figure \ref{fig:thresh}.  From Figure \ref{fig:thresh} we see that the peak intensity curves of the conditioned and recovered samples overlay; with bi-linear fits of each giving lasing thresholds within uncertainty of each other ($E_{T,\text{Cond}}=8.3\pm 1.5$ mJ and $E_{T,\text{Rec}}=7.7\pm1.3$ mJ ).  This suggests that the dye fully recovers, as the lasing threshold is very sensitive to the concentration of undamaged dye molecules \cite{Cao03.02,Yi12.01,Shank75.01,Anderson14.04}.

\begin{figure}
\centering
\includegraphics{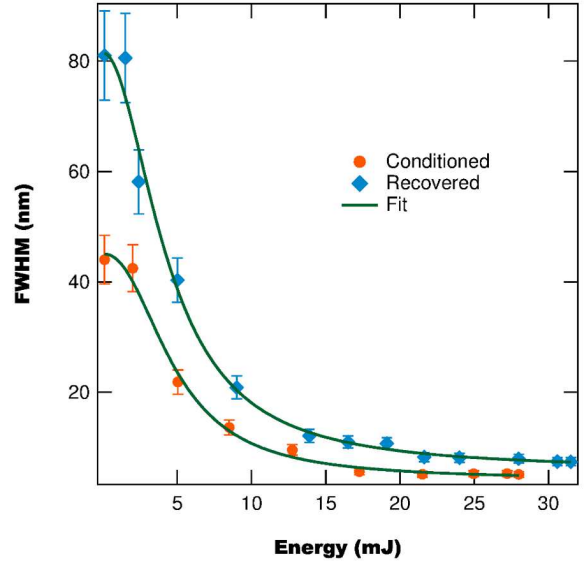}
\caption{FWHM of the sample emission as a function of pump energy for the conditioned sample (squares) and the sample after recovery (circles).  For all pump energies the conditioned FWHM is smaller than the recovered FWHM.  However, the functional shape is similar between the two curves, which is consistent with them having the same lasing threshold.}
\label{fig:widthI}
\end{figure}

Next we measure the emission linewidth as a function of pump energy, shown in Figure \ref{fig:widthI}, for both the conditioned and recovered samples.  From Figure \ref{fig:widthI} we find that while the magnitude of the linewidth differs between the conditioned and recovered samples, the shape of the curves are consistent, with the half-max pump energy of both curves being within uncertainty of each other ($E_{1/2,\text{Cond}}=4.79 \pm 0.60$ mJ and $E_{1/2,\text{Rec}}=4.38 \pm 0.60$ mJ). Once again this result suggests that the lasing threshold is unchanged due to decay and recovery, as the half-max pump energy is related to the lasing threshold \cite{Cao05.01,}. Additionally, Figure \ref{fig:widthI} demonstrates that the recovery process is not reversible for all components of the sample, as the FWHM of the recovered sample is larger than the FWHM of the conditioned sample. If the photodegradation process was completely reversible the linewidth would be unchanged after recovery.

\subsection{Material Characterization}
Up to this point we have focused on the RL properties of R6G+Y$_2$O$_3$/PU during conditioning, decay, and recovery.  However, these results can only go so far in illuminating the underlying mechanisms behind conditioning, decay and recovery.  To gain further insight, we now turn to considering the linear optical properties of the system, namely the visual characteristics and IR-VIS absorbance.

\subsubsection{Images and VIS absorbance.~~}

\begin{figure*}
 \centering
 \begin{subfigure}[b]{0.2\textwidth}
 \includegraphics{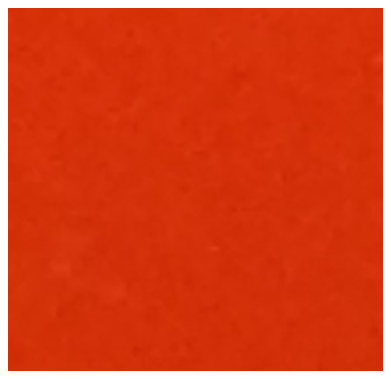}
 \caption[a]{Pristine}
 \end{subfigure}
 \hspace{4em}
\begin{subfigure}[b]{0.2\textwidth}
 \includegraphics{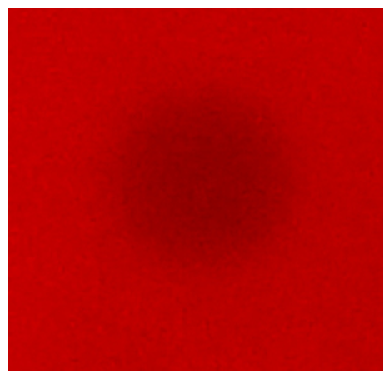}
 \caption[b]{Conditioned}
\end{subfigure}
 \hspace{4em}
\begin{subfigure}[b]{0.2\textwidth}
 \includegraphics{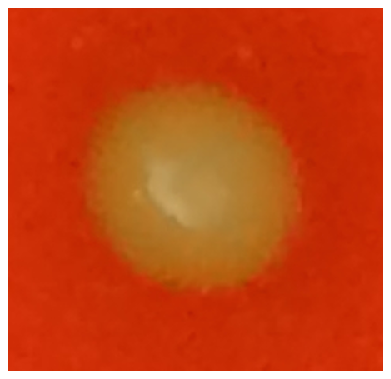}
 \caption[C]{Decayed}
\end{subfigure}
\caption{Images of pristine, conditioned, and damaged RL sample.  The beam conditioned area is found to be darker than the surrounding area, suggesting increased absorption, while after photodegradation the spot is found to have turned yellow, suggesting a decrease in absorption in the green/yellow spectral region. The white area in the decayed spot corresponds to a hot spot in the pump beam.}
\label{fig:picture}
\end{figure*}

To visually chronicle the changes to the sample during irradiation, we first image the pristine sample. We then condition the sample for 5 s, block the pump and image the conditioned area. The pump beam is then unblocked and the sample is irradiated until the sample is sufficiently damaged, at which point it is once again imaged.  Figure \ref{fig:picture} shows the images of the pristine, conditioned, and damaged sample.  From the images we see that conditioning results in a darkening of the sample, while further irradiation results in a yellowing of the damaged area.  The darkening of the conditioned area suggests an increase in the absorption of the sample, while the yellowing of the decayed area suggests a decrease in absorption in the green-yellow region of the spectrum.

%J-Type R6G dimers have a splitting in abs. spectrum leading to decrease in the usual max and an increase in the wings.  Also spec is redshifted....possible trimers as well \cite{Gavrilenko06.01,Arbelo88.01,Bojarski96.01}
%dimers lead to red shifted fluorescence \cite{Bojarski96.01,Penzkofer87.01}

\begin{figure}
 \centering
 \includegraphics{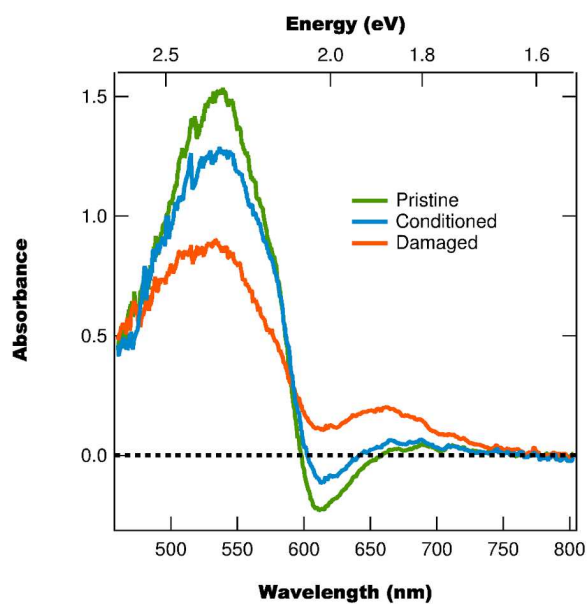}
 \caption{Absorbance of R6G+Y$_2$O$_3$/PU for the pristine, conditioned, and damaged samples.  A negative absorbance corresponds to fluorescent emission from the Rhodamine 6G dye, as the spectra were taken using a broadband white light source.}
 \label{fig:Visabs}
\end{figure}

\begin{figure}
 \centering
 \includegraphics{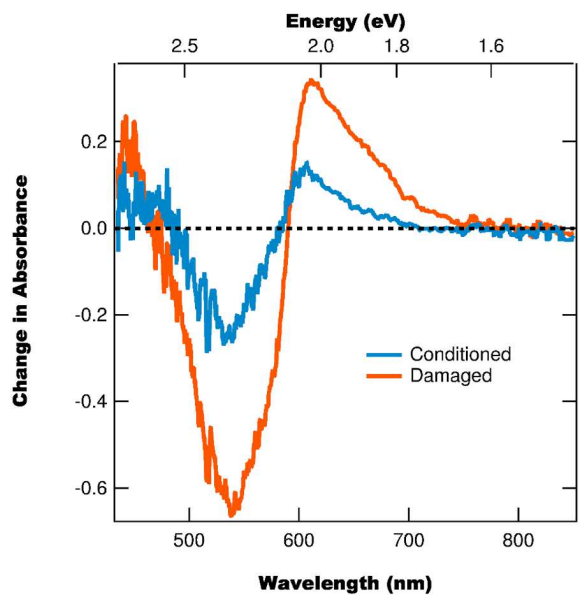}
 \caption{Change in absorbance due to conditioning and degradation.  The two spectra are found to have different spectral behavior with the conditioning curve having narrower peaks and different isosbestic points from the damaged curve.  This suggests different processes are at play within the two regimes.}
 \label{fig:Visdabs}
\end{figure}

To verify these observations we perform surface absorbance measurements on a sample before conditioning, after conditioning, and finally after degradation.  We begin by first measuring the reflected light from a silver mirror to obtain the spectrally flat reflected intensity $I_F(\lambda)$ from the sample plane.  Next we replace the mirror with a sample and measure the spectral intensity of the reflected/emitted light from the sample, $I_S(\lambda)$, which we use to calculate the surface absorbance given by,

\begin{align}
 \text{Abs}(\lambda)=\frac{\rho I_S(\lambda)}{I_F(\lambda)},
\end{align}
where $\rho$ is a factor used to account for the sample and mirror having different baseline reflectances. Figure \ref{fig:Visabs} shows the absorption spectra for the three regimes, with fluorescence included in the absorption calculation, leading to some spectral regions having a negative absorbance value. From Figure \ref{fig:Visabs} we see that as the sample is irradiated the absorption in the green-yellow region (500 nm to 600 nm) decreases, while the absorption in the red and blue spectral regions increases. To better display these changes we plot the change in absorbance (compared to the pristine sample) for the conditioned sample and damaged sample in Figure \ref{fig:Visdabs}.

From Figure \ref{fig:Visabs} and Figure \ref{fig:Visdabs} we observe that as the sample is irradiated the pristine absorbance peak decreases, while two new peaks are formed.  This behavior is different than previously measured for irreversible photobleaching, where the absorbance peak decreases with no new peaks being formed \cite{Kurian02.01,Annieta05.01,George99.01,Ghorai13.01}.  One possible explanation for the observed behavior is the photo-induced formation of J-Type dimers \cite{Gavrilenko06.01,Arbeloa88.01,Bojarski96.01} and/or trimers \cite{Arbeloa88.01}. When R6G forms dimers/trimers its absorbance spectrum redshifts, becomes broader, and has decreased absorbance at the monomer peak \cite{Gavrilenko06.01,Arbeloa88.01,Bojarski96.01}, which is what we observe.  Additionally, the formation of R6G dimers is known to cause the fluorescence spectrum to redshift \cite{Bojarski96.01,Penzkofer87.01}, which can help to explain the observed redshift in the RL emission.

%More dimer papers \cite{Sasai02.01,Matinez04.01,Sasai04.01}

\subsubsection{FTIR Measurements.~~}

\begin{figure}
 \centering
 \includegraphics{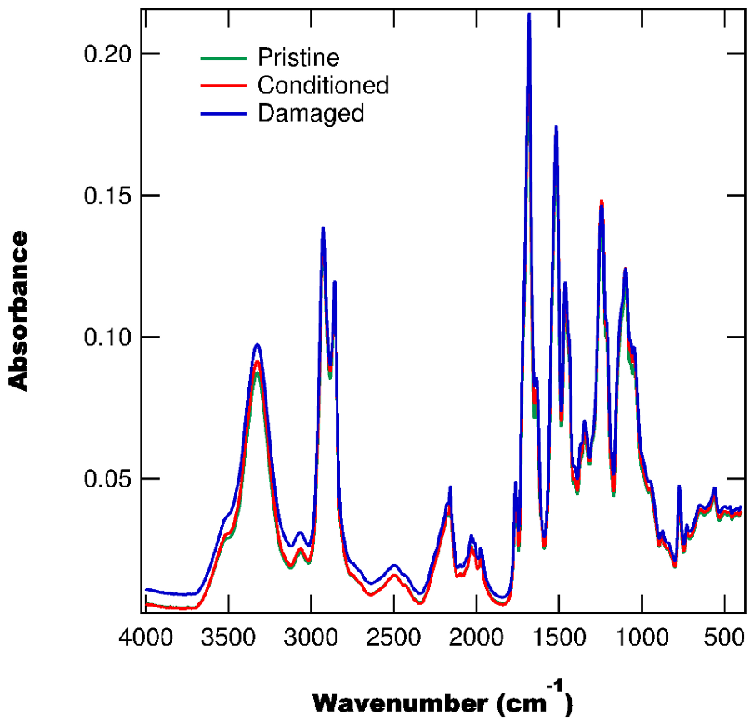}
 \caption{IR Absorbance of the pristine, conditioned and damaged sample.}
 \label{fig:irabs}
\end{figure}

\begin{figure}
 \centering
 \includegraphics{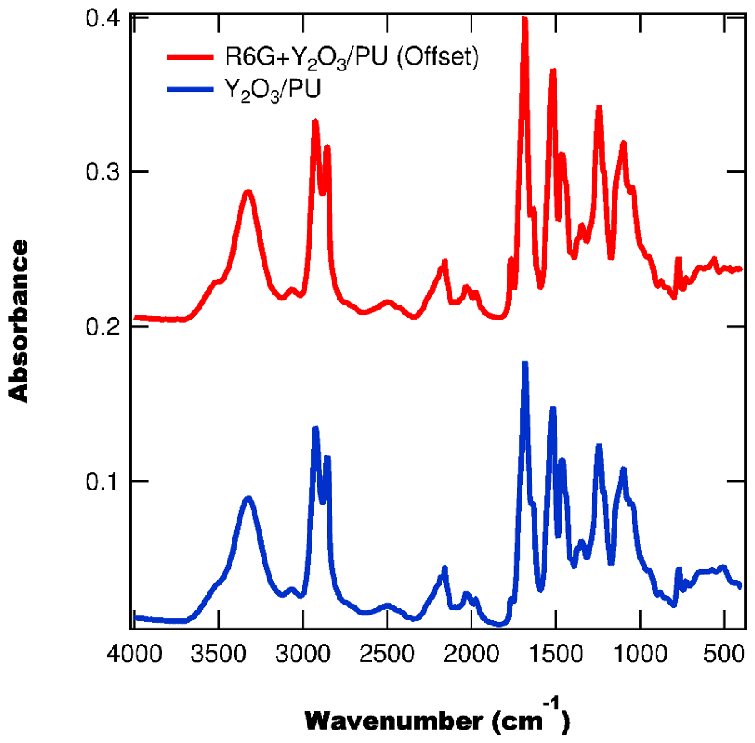}
 \caption{IR absorbance for Y$_2$O$_3$/PU and R6G+Y$_2$O$_3$/PU. The inclusion of the R6G dye results in minimal change to the absorbance spectrum implying that the polymer and NP are the primary contributions to the spectrum.}
 \label{fig:polyabs}
\end{figure}

To identify underlying chemical changes in R6G+Y$_2$O$_3$/PU we use ATR-FTIR to measure the IR absorbance spectrum of the sample before, during, and after irradiation.  Figure \ref{fig:irabs} shows the IR absorbance spectra for the pristine, conditioned, and damaged samples. Given the chemical makeup of R6G, we also measure the IR absorbance spectrum of Y$_2$O$_3$/PU with no dye (shown in Figure \ref{fig:polyabs}) to better determine the source of the IR absorbance peaks.  However, from Figure \ref{fig:polyabs} we find that the IR absorbance spectrum for both nanocomposites (without and with dye) are very similar, with the peak positions consistent between the two, suggesting that Y$_2$O$_3$/PU dominates the IR absorbance.  This result is expected as the amount of dye present is relatively small compared to the amount of polymer and NPs.  Given the lack of an obvious IR absorbance component due to R6G -- with its contributions overwhelmed by the polymer -- it is difficult to say 
definitively that observed changes are due to the dye and/or host.  However, we find that the observed changes are consistent with previous studies of photodegradation in neat PU \cite{Rabek95.01,Nguyen02.01,Wang05.01,Decker99.01,Abuin72.01,Yang70.01,Laue05.01} and therefore we proceed with the assumption that the changes in the IR absorbance spectrum are due to the polymer and not the dye.

\begin{figure}
 \centering
 \includegraphics{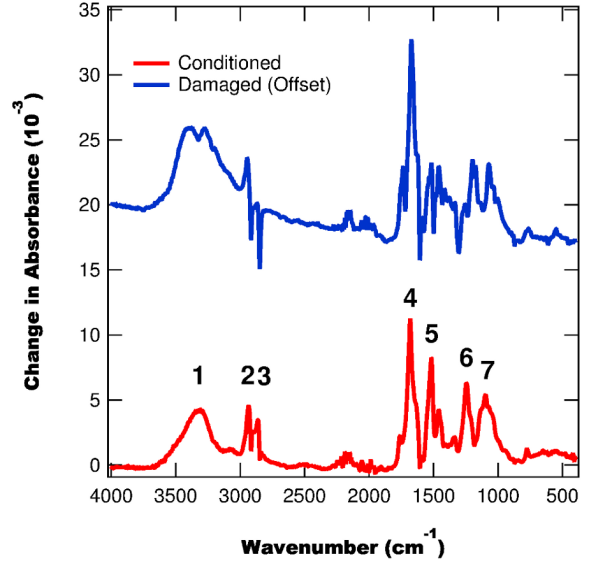}
 \caption{Change in the IR absorbance due to conditioning and degradation. The change in absorbance for the damaged area is offset by 0.015 for clarity.}
 \label{fig:IRdabs}
\end{figure}

Figure \ref{fig:IRdabs} shows the change in IR absorbance of the sample due to conditioning and degradation. From Figure \ref{fig:IRdabs} we find that seven of the IR absorbance peaks from Figure \ref{fig:polyabs} increase due to conditioning and further irradiation results in four of the peaks continuing to increase while the other three decrease.  Table \ref{tab:peaks} lists the peak locations and their assignments. Based on the location and behavior of the peaks -- as well as previous studies on neat polyurethane \cite{Rabek95.01,Nguyen02.01,Wang05.01,Decker99.01} --  we can make several conclusions about the peak behavior:
\begin{enumerate}
 \item The broad peak centered at 3300 cm$^{-1}$ is most likely due to the formation of amines \cite{Rabek95.01,Laue05.01}. 
 \item The increase during conditioning in the peaks at 2944 cm$^{-1}$ and 2858 cm$^{-1}$ correspond to the formation of alkanes arising from the abstraction of $\alpha$-protons from carbonyl fragments \cite{Abuin72.01,Yang70.01,Laue05.01}. The subsequent decrease during further irradiation corresponds to decomposition of the alkanes due either to photooxidation \cite{Calvert08.01} or C-H bond rupture \cite{Mozumder03.01}
 \item The increase of the peak at 1516 cm$^{-1}$ during conditioning corresponds to an increase of N-H bonds, most likely due to photothermally induced curing of leftover monomer \cite{Pretsch00.01}.  The subsequent decrease with further irradiation is consistent with chain scission of polyurethane \cite{Nguyen02.01,Wang05.01}.
 \item The increase in the peak at 1672 cm$^{-1}$ corresponds to the formation of more C=O bonds, which occurs due to photooxidation \cite{Nguyen02.01,Decker99.01}.
 \item The peak at 1244 cm$^{-1}$ is related to the polymerization of the PU and its increase is most likely due to photothermally induced curing of leftover monomer \cite{Pretsch00.01}. 
 \item The increase at 1097 cm$^{-1}$ could possibly be due to the formation of peroxides and hydroperoxides from photooxidation \cite{Pretsch00.01,Rabek95.01,Wicks07.01,Ranby75.01,Cutrone88.01}.
\end{enumerate}

From these observations -- and a review of the literature on photodegradation of neat polyurethanes -- we can conclude that during irradiation the primary mechanism of irreversible polymer photodegradation is Norrish type I photocleavage \cite{Rabek95.01,Laue05.01}, with additional degradation occurring due to photooxidation \cite{Nguyen02.01}. We also conclude that during conditioning leftover monomer undergoes polymerization due to photothermal heating leading to increases in the peaks associated with polyurethane \cite{Pretsch00.01}.  As the neat polymer is transparent at the pump wavelength, the most likely degradation pathway is via energy transfer between the dye and polymer \cite{Moshrefzadeh93.01,Gonzalez00.01,Chang01.01,Rabek95.01} with thermalization \cite{Fellows05.01,Ishchenko08.01} being the largest effect.

\begin{table}[h]
\begin{tabular}{|ccc|}
\hline
\textbf{Peak \#} & \textbf{Center (cm$^{-1}$)} & \textbf{Assignment}     \\ \hline
1                & 3300                 & N-H       \\
2*                & 2944                 & Alkanes                  \\  %Abuin72, Yang70,Laue 05
3*               & 2858                 & Alkanes                  \\
4                & 1672                 & C=O         \\   %tells us about oxidation \cite{Nguyen02.01}
5*               & 1516               & N-H         \\   %tells us about chain scission \cite{Nguyen02.01,Wang05.01}
%6*               & 1302               & C-O-C st asym or C-H st \\
6                & 1244               & N-CO-O \\
7                & 1097               & C-O-C or C-O-O    \\ \hline
\end{tabular}
\caption{Peak locations in the change in IR absorbance from Figure \ref{fig:IRdabs} and their probable assignments. Peaks marked with an asterisk increase during conditioning, and then subsequently decrease with further irradiation.}
\label{tab:peaks}
\end{table}

%\cite{Osawa77.01,Decker91.01}

%\subsection{Summary of results}

\section{Conclusions}
Based on the various measurements we perform on the photodegradation and self-healing of R6G +Y$_2$O$_3$/PU we determine several important results:

\begin{enumerate}
\item When the system is illuminated with intense light its RL emission first increases in amplitude, redshifts, and widens -- in a process we call conditioning -- after which the emission continues to redshift and widen, but now with the amplitude decreasing during decay. 
\item After degradation, the RL emission returns to its initial intensity, but the emission remains redshifted and widened.
\item The net effect of the amplitude fully recovering and the linewidth increasing is an enhancement of the spectrally integrated intensity.
\item Further cycling of the system through degradation and recovery will reach a stable level of enhancement beyond which no further enhancement occurs.
\item Cycling is found to decrease the photostability of the system, while the recovery rate and efficiency is maintained.
\item Degradation results in the pristine absorbance peak decreasing, while a new broadened and redshifted peak forms, which is consistent with the formation of R6G dimers and trimers.
\item Changes in the FTIR spectrum suggest that both Norrish type I photocleavage and photooxidation are responsible for the irreversible photodegradation of the polymer host.
\end{enumerate}

From these results we conclude that there are at least two species of damaged products formed during photodegradation: one reversibly damaged and the other one irreversibly damaged.  Given the irreversible changes in the RL linewidth and peak location, as well as the irreversible changes in the IR-VIS absorbance, we can conclude that the irreversibly damaged species is primarily related to Norrish type I photocleavage and photooxidation of the polymer host, with additional changes due to the photoinduced formation of R6G dimers and trimers.  Since the polymer host is transparent at the pump wavelength, its damage is most likely mediated by energy transfer between the dye and polymer \cite{Moshrefzadeh93.01,Gonzalez00.01,Chang01.01,Annieta,Rabek95.01,Fellows05.01,Ishchenko08.01}. To further refine the identification of the irreversible damaged species we plan on performing dose-dependent studies using imaging \cite{Anderson11.01}, absorbance \cite{embaye08.01,Anderson14.02,Ramini11.01}, interferometry \cite{Anderson13.02,Anderson14.03} and FTIR.

As to the nature of the reversible portion of photodegradation, our current results remain inconclusive.  However, the underlying mechanism is most likely similar to that proposed for DO11/PMMA. In that system, dye molecules are hypothesized to form domains of correlated molecules along polymer chains \cite{Ramini13.01,Anderson14.01} with reversible degradation resulting in the formation of charged fragments \cite{Desau09.01,Anderson14.01}.  These fragments are confined in space by the polymer and recovery occurs when molecular interactions within a domain result in the recombination of the fragments \cite{Anderson14.01}.  To determine if charged species are formed, and possibly identify them, we are planning degradation and recovery studies with an applied electric field \cite{Anderson13.01}.

\section{Acknowledgements}
This work was supported by the Defense Threat Reduction Agency, Award \# HDTRA1-13-1-0050 to Washington State University.

\section{Compliance with Ethical Standards}
Conflict of Interest: The authors declare that they have no conflict of interest.

% \bibliographystyle{osajnl}
% \bibliography{PrimaryDatabase,ASLbib} 

\end{document}